\documentclass{ws-ijmpd}

\newcommand{\f}{\mathbf f}

\newcommand{\T}{\mathbf T}

\begin{document}

\markboth{Efra\'\i n Rojas}
{Higher order curvature terms in BI type brane theories}

%
%

\title{HIGHER ORDER CURVATURE TERMS IN BORN-INFELD TYPE BRANE THEORIES}

\author{EFRAIN ROJAS}

\address{Departamento de F\'\i sica, Facultad de F\'\i sica e Inteligencia
Artificial, Universidad Veracruzana\\
Xalapa, Veracruz, 91000, M\'exico
\\
efrojas@uv.mx}

\maketitle


\begin{abstract}
The field equations associated to Born-Infeld type
brane theories are studied by using an auxiliary variables method.
This approach hinges on the fact that the expressions defining the
physical and geometrical quantities describing the worldvolume are 
varied independently.
The general structure of the Born-Infeld type theories for branes
contains the square root of a determinant of a combined matrix 
between the induced metric on the worldvolume swept out by the 
brane and a symmetric/antisymmetric tensor depending of gauge, 
matter or extrinsic curvature terms taking place on the worldvolume. 
The higher order curvature terms appearing in the determinant form come to 
play as competition with other effective  brane models. 
Additionally, we suggest a Born-Infeld-Einstein type action for branes 
where the higher order curvature content is provided by the 
worldvolume Ricci tensor. This action provide an alternative 
 description of the dynamics of braneworld scenarios.
\end{abstract}


\section{Introduction}

The Born-Infeld (BI) theory was originally designed to overcome
the infinity problem of a point charge source in the standard Maxwell
electromagnetism\cite{Born,Born1}. Over the years this idea was 
almost forgotten until string theory revealed the existence of certain class 
of extended objects named $Dp$-branes which brought altogether a revival 
of the original BI proposal\cite{Polchinski1,clifford}. These extended objects 
provide explicit realizations of several interesting phenomena in a wide 
range of physical theories at the tiniest scales, mainly in modern 
non-perturbative string theory. The simplest action governing the dynamics of 
$Dp$-branes is the well known Dirac-Born-Infeld (DBI) action which contains the square
root of the determinant of a metric constructed with the induced
metric on the worldvolume swept out by the $Dp$-brane and the 
electromagnetic field strenght. This action has given rise to an intense 
research and we have witnessed a growing development 
of the subject which has triggered a variety of suggestions
akin to it. A related geometrical construction 
ignoring electromagnetism altogether but instead including higher 
order curvature terms in string theory was proposed long time ago by Lindstrom 
{\it et al}\cite{Lindstrom2,Lindstrom}. There, such structure involves the 
determinant of a metric which is now the sum of the induced metric and a symmetric
tensor built from the extrinsic curvature of the worldsheet swept out
by the string. This effective BI type model suffers however severe drawbacks mainly due to
its second order derivative dependance and in consequence has not
received much attention. Needless to say, these are elegant gravitational 
theories defined on surfaces, with a 
determinant Lagrangian density form, which 
still deserve a careful analysis because of their attractive geometrical properties.
This type of theories adapted to branes is the subject we will focus on this work.

As is by now well known in the brane context, despite its many successes, 
there are many contexts where the Dirac-Nambu-Goto (DNG) action
is not adequate and it is natural to consider models that depend on
higher derivatives of the embedding variables through the extrinsic curvature
of the worldvolume swept out by the brane. 
The most general field theoretical effective action governing the dynamics of a brane results in an expansion
constructed out of the geometrical scalars of its worldvolume\cite{brandon-ruth,anderson}. 
These extrinsic curvature
correction terms have found extensive use in concrete applications. For instance,
the addition of a rigidity term\footnote{The criterion for rigidity that we will
follow in this paper is that
 is as a correction to DNG objects in order to penalize singularities
in their evolution and thus favouring a variety of configurations richer
in geometrical and physical structure
than DNG objects. On physical grounds, the inclusion of this type of correction 
terms simply means that rigid branes resist being bent.
} in an effective action for stringy 
QCD\cite{polyakov,kleinert}, the systematic approximation schemes that 
arise in expansions in the thickness of topological defects\cite{ruth1,ruth2,ruth3}, 
or in actions  appearing in the braneworld scenarios\cite{randall,cadoni}, to mention
some.
Therefore, there is no field theoretical reason 
that does not preclude the possibility
of consider an alternative expansion but, enclosed now in a 
geometrical BI type action solely, without introducing 
electromagnetism.

The mechanical content of an extended object is captured by its 
geometrical degrees of freedom where the conserved 
stress tensor plays a very important role. This determines 
the dynamics of the extended object\cite{Noether}
thus providing a very powerful geometrical tool for the description
of deformations of branes. On physical
grounds such stress tensor is merely the momentum density 
of the brane. On the one hand, the dependance on the independent variables 
of the effective model governing the evolution of the brane is the 
first step to be recognized in order to perform a variational process. 
One then proceed in the usual way to obtain the dynamical laws 
that the system must obey. On the other hand, the standard variational 
process for a second order field theory is a non-trivial 
task which gives rise to annoying computations in order to derive equations 
of motion and it is sharply elaborated for determinant theories. 
Certainly this unpleasant fact appears to be a great difficulty but
this is not the road we take here. To bypass this field theoretical
problem we follow an original 
strategy for reaching geometrical and physical information, based in
an auxiliary variables method (AVM) introduced by Guven in order to describe 
fluctuating surfaces\cite{Guven}. The cornerstone in this approach consists 
of promoting inherent geometrical quantities, defined  on the worldvolume, 
to play the role of auxiliary variables, distributing the original 
deformation of the embedding variables among the auxiliary variables. 
The response of the action to a deformation of its worldvolume reflects 
in the conservation of the stress tensor.

In this paper, equipped with modern variational techniques, we provide 
the dynamical information supplied
by the conserved stress tensor for BI type brane theories.  
We examine mainly second order brane theories that depend on the 
extrinsic curvature, which possess a determinant structure. 
We would like to point out that, with a suitable choice of constraints,  
the conserved stress tensor and the field equations are obtained effortless
through the use of the AVM, elucidating also the geometrical nature of BI
type brane models. The paper is structured 
as follows. In Section 2, in order to gain insight into the 
AVM for BI objects, we adapt it to study
the well known $Dp$-brane dynamics. We obtain with no effort the mechanical 
content of $Dp$-branes\cite{dbigeometry}. We thus pave the way to the application 
of the AVM to more complex BI type theories. In Section 
3 we follow closely the AVM introduced by Guven\cite{Guven} in order to study the geometrical 
properties of a rigid BI action for branes, originally proposed in the 
string theory context\cite{Lindstrom2,Lindstrom}. By contrast, the resulting equations of motion 
are highly non-trivial. In Section 4  we introduce an action in the brane 
context that mimics the Born-Infeld-Einstein (BIE) suggestion by Deser and 
Gibbons to modify Einstein gravity\cite{deser}. 
We suggest that an alternative
approximation for the study of brane world scenarios lie in BIE type models.
This action encodes
a rich geometrical content which results promising to explore cosmological
brane models which deserves a careful analysis.
We conclude in Section 5 with some comments and we discuss 
briefly our findings. To make our work selfcontained, we develop
the general auxiliary variables method for a general relativistic extended object
whose action include extrinsic curvature. This topic will be developed in
\ref{app:AVM}. Important mathematical expressions involving the determinant
of a matrix have been collected in \ref{app:mathidentities}.

\section{DBI objects}

Consider a $Dp$-brane, or in general a relativistic extended object 
denoted by $\Sigma$, of dimension $p$ evolving in a $N$-dimensional 
background spacetime endowed with an arbitrary metric $G_{\mu \nu}$\,\, 
$(\mu, \nu = 0,1,2, \ldots, N-1)$. The trajectory, or worldvolume 
$m$ swept out by $\Sigma$ is an oriented timelike manifold of 
dimension $p+1$, described by the embedding functions $x^\mu = 
X^\mu(\xi^a)$ where $x^\mu$ are local coordinates of the background 
spacetime, $\xi^a$ are local coordinates of $m$, and $X^\mu$ are 
the embedding functions $(a,b= 0,1,2,\ldots,p)$. The metric induced 
on the worldvolume from the background is given by 
\begin{equation}
g_{ab}= G_{\mu \nu}e^\mu
{}_a e^{\nu}{}_b := e_a \cdot e_b\,,
\label{eq:gab}
\end{equation}
where  $e^\mu {}_a = \partial_a X^\mu$ are the tangent vectors to $m$.
Here and henceforth a central dot indicates contraction using the
background metric.
In this framework we introduce $N-p-1$ normal vectors to the 
worldvolume, denoted by $n^\mu {}_i \,\,(i=1,2,\ldots,N-p-1)$. 
These are defined implicitly by $n\cdot e_a = 0$ and $n_i \cdot n_j = 
\delta_{ij}$ where $\delta_{ij}$ is the Kronecker delta symbol.

The well known DBI action governing the low energy dynamics of 
$Dp$-branes is 
\begin{equation}
S_{{\mbox{\tiny DBI}}}[X,A]=  \beta_p \int_m d^{p+1}\xi \,
\sqrt{-{\mbox{det}} (g_{ab} + {\cal F}_{ab})},
\label{eq:DBIaction}
\end{equation}
where $\beta_p$ is the tension of the $Dp$-brane,
${\cal F}_{ab} = \alpha\,F_{ab} + B_{ab}$ is a gauge invariant 
quantity defined in terms of
\begin{equation} 
F_{ab} =2\partial_{[a}A_{b]}  \qquad {\mbox{and}} \qquad B_{ab}= 
B_{\mu \nu}e^\mu {}_a e^\nu{}_b\,,
\label{eq:Fab}
\end{equation}
where $F_{ab}$ is the electromagnetic field strength associated to a $U(1)$ 
gauge field $A_a$ living on $m$ and $B_{ab}$ is the pullback to the 
worldvolume of the Neveu-Schwarz (NS) 2-form $B_{\mu \nu}$ and $\alpha$ is the 
BI parameter related to the inverse tension of branes\cite{Polchinski1,clifford}.
Among the important features of the action~(\ref{eq:DBIaction}) 
to be mentioned are that it is a first order derivative theory, its 
invariance under worldvolume reparametrizations and its invariance 
under a NS gauge transformation, $B_{ab} \to B_{ab} - 2\partial_{[a} 
\lambda_{b]}$ if we shift the $U(1)$ field $A_a \to A_a + \alpha^{-1} 
\lambda_a$ where $\lambda_a$ is a $1-$form; in other words, ${\cal F}_{ab}$ 
is the gauge invariant quantity in the presence of NS background field. 

The response of the action~(\ref{eq:DBIaction}) to a deformation of the 
surface $X \to X + \delta X$, as well as to a deformation of the $U(1)$ 
gauge field $A \to A + \delta A$ turns out to be a rather involved 
computation\cite{dbigeometry}. The source of this difficulty come 
from the definitions of the tensors $g_{ab}$ and ${\cal F}_{ab}$ which 
encode  the derivatives of the field variables via the structural 
relationships~(\ref{eq:gab}) and~(\ref{eq:Fab}) and also due to the fact that 
the background fields might depend of the embedding variables as occurs 
in more interesting $Dp$-brane scenarios. 

To circumvent the usual variational procedure our strategy will be to adopt the 
action~(\ref{eq:DBIaction}) as a functional of the independent variables 
$g_{ab}$ and ${\cal F}_{ab}$ instead of the usual ones, 
$X^\mu$ and $A_a$ through the Lagrangian density ${\cal L}_{\mbox{\tiny DBI}}
= {\cal L}_{\mbox{\tiny DBI}} (g_{ab},{\cal F}_{ab})$. We thus construct the new functional action
\begin{eqnarray}
S&& \!\!\!\!\![  X,e{}_a,g_{ab},{\cal F}_{ab},{\cal F}^a,\mathfrak{J}^{ab},{\cal T}^{ab}] =  
S_{\mbox{\tiny DBI}}[g_{ab},{\cal F}_{ab}] 
+ \int_m dV\,{\cal F}^a \cdot (e {}_a - \partial_a X)  \nonumber \\
&-& \frac{1}{2} \int_m dV {\cal T}^{ab}(g_{ab} - e{}_a \cdot e{}_b) 
- \frac{1}{2} \int_m dV
{\cal J}^{ab} \left( {\cal F}_{ab} - 2 \alpha \partial_{[a}A_{b]}
- B_{\mu \nu}e^\mu {}_a e^\nu{}_b\right), 
\label{eq:total-act}
\end{eqnarray}
where $dV =\sqrt{-g}\,d^{p+1}\xi$ is the worldvolume element. 
${\mathcal F}^a, {\mathcal J}^{ab}$ and ${\mathcal T}^{ab}$ are
Lagrange multipliers.
Note that ${\cal T}^{ab}$ and 
${\cal J}^{ab}$ are symmetric and antisymmetric, respectively. 
To deduce the dynamics we will take a shorcut treating $e_a,g_{ab}
,{\cal F}_{ab}$ and $X$ as independent auxiliary variables.  
The implementation of an AVM to study another type of bosonic
brane theories is described in \ref{app:AVM}.

The variation of the total action~(\ref{eq:total-act}) with respect
to the embedding functions manifiests as 
\begin{equation}
\nabla_a {\cal F}_\mu{}^a  = -\frac{1}{2}\left( 
{\cal T}^{ab}\partial_\mu G_{\alpha \beta} + {\cal J}^{ab} 
\partial_\mu B_{\alpha \beta}\right)e^\alpha{}_ae^\beta{}_b \,,
\label{eq:var1} 
\end{equation}
where $\nabla_a$ is the covariant derivative compatible with $g_{ab}$. 
Now, the Euler-Lagrange (EL) derivative with
respect to the tangent vectors $e{}_a$ is
\begin{equation}
{\cal F}_\mu{}^a= -\left( {\cal T}^{ab} G_{\mu \nu} + {\cal J}^{ab}
B_{\mu \nu}\right) e^\nu{}_b \,,
\label{eq:f1}
\end{equation}
which results tangential to the worldvolume. To determine the form of the 
worldvolume stress tensor~(\ref{eq:f1}) it is necessary to know completely 
${\cal T}^{ab}$ and $ {\cal J}^{ab}$. Varying over the induced metric we get 
\begin{equation}
{\cal T}^{ab}= \frac{2}{\sqrt{-g}}\left( \frac{\partial 
{\cal L}_{\mbox{\tiny DBI}}}{\partial g_{ab}}\right) = 
\beta_p \frac{\sqrt{-{\cal M}}}{\sqrt{-g}}\left( {\cal M}^{-1}\right)^{(ab)} \,,
\label{eq:lab}
\end{equation}
which is nothing but the metric stress tensor associated to the DBI
action. Here,  $({\cal M}^{-1})^{ab}$ denotes the inverse matrix of ${\cal M}_{ab}:=
g_{ab} + {\cal F}_{ab}$, such
that $({\cal M}^{-1})^{ac}{\cal M}_{cb} = \delta^a{}_b$ and 
${\cal M}= {\mbox{det} ({\cal M}_{ab})}$. Further, $\left( {\cal M}^{-1}\right)^{(ab)}$
and  $\left( {\cal M}^{-1}\right)^{[ab]}$ denote the symmetric and the
antisymmetric parts of the matrix $\left( {\cal M}^{-1}\right)$, respectively\footnote{The 
symmetrization and the antisymmetrization notation 
over the indexes of tensors is the standard one, $S_{(ab)} = \frac{1}{2}(S_{ab} 
+ S_{ba})$ and $S_{[ab]} = \frac{1}{2}(S_{ab} - S_{ba})$.
}.
Now, with respect to the $U(1)$ field dependance of~(\ref{eq:total-act}), we first 
compute the EL derivative with respect to $A$,
\begin{equation}
\nabla_b {\cal J}^{ba}=0\,.
\label{eq:var2} 
\end{equation}
This reflects into a conservation law. The remaining
EL derivative with respect to ${\cal F}_{ab}$ reads
\begin{equation}
{\cal J}^{ab} = \frac{2}{\sqrt{-g}}\left( \frac{\partial 
{\cal L}_{\mbox{\tiny DBI}}}{\partial {\cal F}_{ab}}\right) =
- \beta_p \frac{\sqrt{-{\cal M}}}{\sqrt{-g}}
\left( {\cal M}^{-1}\right)^{[ab]}\,.
\label{eq:fab}
\end{equation}
This complements the geometrical information to determine the 
stress tensor~(\ref{eq:f1}). ${\cal J}^{ab}$ is the excitation 
tensor on the worldvolume\cite{clifford,dbigeometry}.  
Thus, the stress tensor~(\ref{eq:f1}) is written in terms of geometrical
and physical tensors. 

By means of a straightforward computation, the divergence expression ~(\ref{eq:var1}) yields 
\begin{equation}
\nabla_a {\cal T}^{ab}\,e_{\mu \,b} - {\cal T}^{ab}K_{ab} ^i\,n_{\mu\,i} = -\frac{1}{2}
{\cal J}^{ab}\left( \partial_\mu B_{\beta \alpha} +\partial_\beta B_{\alpha \mu} +
\partial_\alpha B_{\mu \beta} \right)e^\alpha{}_a e^\beta{}_b \,, 
\label{eq:almost}
\end{equation}
where we have exploited the Gauss-Weingarten equations,
$\nabla_a e^\mu{}_b = -K_{ab} ^i n^\mu{}_i -\Gamma^\mu _{\alpha \beta}
e^\alpha{}_a e^\beta{}_b$, where $K_{ab} ^i$ is the extrinsic curvature 
of the worldvolume $m$\cite{Spivak,defo} and $\Gamma^\mu _{\alpha \beta}$
are the connection coeficients associated to $G_{\mu \nu}$. The worldvolume 
projections of the relation~(\ref{eq:almost}) are
\begin{eqnarray}
{\cal T}^{ab}K_{ab} ^i &= {\cal F}^i\,,
\label{eq:1moto}
\\
\nabla_a {\cal T}^{ab} &=0\,,
\label{eq:0moto}
\end{eqnarray}
where ${\cal F}^i = \frac{1}{2}
{\cal J}^{ab}H_{\alpha \beta \mu}
e^\alpha{}_a e^\beta{}_b n^{\mu \,i}$ with $H_{\alpha \beta \mu} = 
\partial_\mu B_{\beta \alpha } + \partial_\alpha B_{\mu \beta} + 
\partial_\beta B_{\alpha \mu}$ being the NS strength 3-form field which 
satisfies $dH = 0$. We can recognize immediately the equations of motion 
(\ref{eq:1moto}) as those appearing by using the original variables 
$X$\cite{dbigeometry}. The equations~(\ref{eq:0moto}) are consistency 
conditions which reduce to geometrical identitites\cite{Noether}. These 
equations are accompanied with the conservation law for 
the bicurrent ${\cal J}^{ab}$.

Note therefore that we have $N-p-1$ equations of motion for $X$,~(\ref{eq:1moto}), 
and $p+1$ equations of motion for $A$,~(\ref{eq:var2}). 
The consistency conditions~(\ref{eq:0moto}) are consequence of the 
reparametrization invariance of the action $S_{\mbox{\tiny DBI}}$.
Following Carter\cite{Carter2}, it is worthy of mention that~(\ref{eq:1moto}) resembles Newton's 
second law where ${\cal T}^{ab}$ plays the role of a mass, $K_{ab} ^i$
the generalization of the acceleration and ${\cal F}^i$ may roughly
be viewed as a force density.
In the same spirit, the equations~(\ref{eq:var2}) yield Maxwell
equations with support on the worldvolume. For completeness in our 
geometrical description we would like to mention that the following identity holds
\begin{equation}
{\cal T}^a {}_c = (-g)^{-1/2}{\cal L}_{\mbox{\tiny{DBI}}} \delta^a{}_c
- \alpha^{-1}{\cal J}^{ab}{\cal F}_{bc}\,.
\label{eq:id1}
\end{equation}

\section{Born-Infeld-Lindstrom-Ro\v{c}ek-van Nieuwenhuizen branes}

Instead of adding certain matter terms to the worldvolume metric, like scalar 
fields, into the combined BI matrix we can consider another tensors
of different nature. In this section we study a more complex BI type 
theory for branes, originally proposed by Lindstrom {\it et al} in order
to describe, in a first-order approach, a Weyl invariant 
string\cite{Lindstrom2,Lindstrom}. 
The main feature of this theory is the inclusion of the extrinsic
curvature of the worldsheet. We put forward for consideration such 
suggestion for extended objects of arbitrary dimension which evoke our 
interest in its geometrical properties. Consider the action 
\begin{equation}
 S_{{\mbox{\tiny{BILRvN}}}}[X] =  a \int_m d^{p+1}\xi\,
\sqrt{-{\mbox{det}(g_{ab} + b\,f_{ab})}}\,,
\label{eq:action}
\end{equation}
where 
\begin{equation}
f_{ab}=K_a {}^{c\,i}K_{cb\,i},
\end{equation}
is a worldvolume symmetric tensor, 
$a$ being the tension of the extended object and $b$ is the concomitant
dimensional constant 
characterizing the relative weight of non-linear
terms. Contrary to the DBI case, the extended metric $M_{ab} := 
g_{ab} + bf_{ab}$ becomes symmetric. We restrict ourselves in the rest 
of the work to consider a flat Minkowski as background spacetime for 
the sake of simplicity. Thus, $G_{\mu \nu} = \eta_{\mu \nu}$. 

The action~(\ref{eq:action}) will be referred hereafter to as 
Born-Infeld-Lindstrom-Ro\v{c}ek-van Nieuwenhuizen 
action (BILRvN) for branes. The symmetry underlaying this action is the invariance
under worldvolume reparametrizations.  
Because of the explicit extrinsic curvature dependence for $f_{ab}$, 
the BILRvN action when expanded, falls into the collection of second order  
actions for strings or membranes\cite{Noether}. 
The situation here is slightly different in comparison with the developed
one for the case of $Dp$-branes. Now, both the tensors $g_{ab}$ and $K_{ab} ^i$
encode derivatives of the original variables $X$. 
Bearing in mind the way followed for the $Dp$-brane case now we shall consider~(\ref{eq:action}) 
as an action functional of the variables $g_{ab}$ and $K_{ab} ^i$ through the 
Lagrangian density ${\cal L}_{\mbox{\tiny BILRvN}} = {\cal L}_{\mbox{\tiny 
BILRvN}}(g_{ab},K_{ab} ^i)$. We concern now to compute the response of the action~(\ref{eq:action}) 
to a deformation of the worldvolume. The convenient strategy will 
be to distribute the deformation $X \to X + \delta X$ among independent 
auxiliary variables. In \ref{app:AVM} we have developed the general
framework to describe actions involving extrinsic curvature by invoking an AVM.
Here we shall use those general results.

For our present case, by using the results (\ref{eq:Tab}) and (\ref{eq:Labi}),
an straightforward computation leads to
\begin{eqnarray}
T^{ab}&=& \frac{2}{\sqrt{-g}} \left( \frac{\partial {\cal L}_{\mbox{\tiny 
BILRvN}}}{\partial g_{ab}}\right)  = a\frac{\sqrt{-M}}{\sqrt{-g}}(M^{-1})^{cd}
[\delta_c{}^a \delta_d{}^b
- b\, K_{c}{}^{a\,i}K^b{}_{d\,i}].
\label{eq:Lambda}
\\
\Lambda^{ab} _i &=& - \frac{1}{\sqrt{-g}} \left( \frac{\partial {\cal L}_{\mbox{\tiny 
BILRvN}}}{\partial K_{ab} ^i}\right)  = -ab\frac{\sqrt{-M}}{\sqrt{-g}}(M^{-1})^{c(a}K^{b)}{}_{c\,i}\,.
\label{eq:KK}
\end{eqnarray}
where $ (M^{-1})^{ab}$ 
stands for the inverse matrix of $M_{ab}$ such that $(M^{-1})^{ac}M_{cb} 
= \delta^a {}_b$ and $M= {\mbox{det}}(M_{ab})$. 
We would like to remark that $T^{ab}$ do correspond to the metric stress tensor.

From Eq. (\ref{eq:f-final}), the conserved stress tensor then acquires the form
{\small
\begin{equation}
 f^{a}= - \left( T^{ab} + ab \frac{\sqrt{-M}}{\sqrt{-g}}
 (M^{-1})^{c(a}K^{d)}{}_{c\,i}  K_d{}^{b\,i} \right)  \,e_b 
 - \frac{ab}{\sqrt{-g}} \widetilde{\nabla}_b \left[ \sqrt{-M}
 (M^{-1})^{c(a}K^{b)}{}_{c\,i} \right]   \,n^{i}\,,
\label{eq:f12} 
\end{equation}}
where $\widetilde{\nabla}_a$ is the $O(N-p-1)$ covariant derivative
on $m$ and also invariant under normal rotations \cite{defo}. 
To determine the equations of motion as well the geometrical 
consistency conditions associated to this BI type action, we project the 
conservation law~(\ref{eq:CC}), taking into account (\ref{eq:f12}), 
along the worldvolume basis. Whereas
the part proportional to $n^{\mu\,i}$ implies the equations of motion
\begin{equation}
T^{ab}K_{ab} ^i
= F^i\,,
\label{eq:Moto}
\end{equation}
with
\begin{equation*}
\sqrt{-g}F^i = ab \widetilde{\nabla}_a\widetilde{\nabla}_b \left[ 
\sqrt{-M} (M^{-1})^{c(a}K^{b)}{}_{c\,i}\right] 
-ab \sqrt{-M} (M^{-1})^{d(a}K^{c)}{}_{d\,j}K_c{}^{b\,j}K_{ab} ^i\,,
\end{equation*}
the tangential part becomes geometrical consistency conditions 
(see (\ref{eq:tangential})).
The equations of motion (\ref{eq:Moto})
are the normal projections of the conservation law
(\ref{eq:CC}). Only the normal deformations are physical whereas
the tangential deformations simply reduce to geometrical identities
associated to reparametrizations of the worldvolume coordinates.
Note that we have $N-p-1$ equations of motion, one for each normal. 
Despite that the relation (\ref{eq:Moto}) resembles Newton's second law, 
as menntioned in previous section, this is not the case. The pitfall
in this reasoning
is that (\ref{eq:Moto}) comprises a set of $N-p-1$ partial differential
equations of fourth order in derivatives of $X$. 
Explicitly, the equations of motion (\ref{eq:Moto}) are given by
\begin{eqnarray}
\sqrt{-M}\left[ g^{ab} + \left( {M^{-1}}\right)^{ab} - b \left( 
{M^{-1}}\right)^{cd}
K_c{}^{a\,j}{}K^b{}_{d\,j}  \right]&& K_{ab} ^i 
\nonumber
\\ 
&&\!\!\!\!\!\!\!\!\!\!\!\!\!\!\!\!\!\!\!\!\!\!\!\!\!\!\!\!= 2b\,\widetilde{\nabla}_a \widetilde{\nabla}_b \left[ \sqrt{-M}\left( 
{M^{-1}}\right)^{c(a} K_{c}{} ^{b)\,i}  \right],
\label{eq:eom-r}
\end{eqnarray}
where we have used the expressions (\ref{eq:Lambda}) and (\ref{eq:KK}). 
These equations of motion are in agreement with those equations of motion
obtained by usual variation methods\cite{Noether,defo}
The form of these equations of motion are neat and compact in comparison
with the original ones \cite{Lindstrom2}.
Further,
for this BILRvN theory, by defining $\sqrt{-g}J^{ab} := a\,\sqrt{-M}
\left( M^{-1}\right)^{cd} K^a {}_{c\,i} K_d {}^{b\,i}$, the following 
identity holds
\begin{equation}
T^{ab}M_{bc} = (-g)^{-1/2} {\cal L}_{\mbox{\tiny{BILRvN}}}\delta^a{}_c
- b\,J^{ab}M_{bc}\,,
\label{eq:id2}
\end{equation}
which looks similar to (\ref{eq:id1}).

\subsection{Alternative geometry}

We can adopt the viewpoint of an alternative volume 
element depending of the worldvolume metric and the term quadratic in the
extrinsic curvature given by $f_{ab}$. By allowing for an extended metric, $M_{ab} =
g_{ab} + b f_{ab}$, we go beyond the ordinary theory. Suppose now that ${\cal D}_a$
is the covariant derivative compatible with $M_{ab}$. Then, according to this
new geometry the corresponding connection coefficients, $C^a _{bc}$, can be computed 
straightforwardly\cite{wald}
\begin{eqnarray}
C^a _{bc} &=& \Gamma^a _{bc} + \frac{b}{2} (M^{-1})^{ad} (\nabla_b f_{cd}
+ \nabla_c f_{bd} - \nabla_d f_{bc})
\nonumber
\\
&=&  \Gamma^a _{bc} + b\, (M^{-1})^{ad} 
{\cal E}_{d\,i} \cdot \widetilde{\nabla}_b {\cal E}_c {}^i\,,
\label{eq:Cs}
\end{eqnarray}
where we have introduced the shorthand notation ${\cal E}_a {}^i := 
\widetilde{\nabla}_a n^i$ and $\Gamma^a _{bc}$ are the connection 
coefficients compatible with the induced metric $g_{ab}$. 
Bearing in mind that $\Gamma^a _{bc} = g^{ad} e_d \cdot \partial_b e_c$
we note that ${\cal E}_a {}^i$ plays the role of tangent vectors.
The corresponding relationship between the Ricci tensors is the following
\begin{equation}
{\cal R}_{ab} = {R}_{ab} + {\cal D}_a {S}^c _{bc} - {\cal D}_c 
{S}^c _{ab} - {S}^d _{ac} {S}^c _{bd} + 
{S}^d _{ab} {S}^c _{cd}\,,
\end{equation}
where $R_{ab}$ is the Ricci tensor associated with the connection
$C^a _{bc}$ and we have defined the tensor field
\begin{equation}
 {S}^a _{bc} := b \,({M}^{-1})^{ad} {\cal E}_{d\,i} \cdot \widetilde{\nabla}_b 
{\cal E}_c {}^i\,.
\end{equation}

In closing this subsection we expand the integrand of the action~(\ref{eq:action}) 
using the expression~(\ref{eq:expansion}). It is straightforward to note
that
{\small
\begin{equation}
\sqrt{-M} = \sqrt{-g} \left\lbrace 1 + \frac{b}{2}\left( K^i K_i -
{\cal R} \right) + \frac{b^2}{8} \left[ \left( K_{ab} ^i K^{ab} _i \right)^2 
- 2 K^{ab} _i K_{bc} ^i K^{cd} _j K_{da} ^j \right] + {\cal O}(K^3) \right\rbrace 
\end{equation}}
where we have used the contracted Gauss-Codazzi condition,
${\cal R}=K^iK_i - K_{ab} ^i K^{ab} _i$ \cite{defo}.
We observe that this expansion produces a vast array of interactions. 
To first order in
$b$ we have a mixture of the Dirac-Nambu-Goto model (DNG), the Polyakov action\cite{polyakov}
and the Regge-Teitelboim (RT) model\cite{RT,ostro}. The case of a string ($p=1$)
is quite special since the integral of $\sqrt{-g}{\cal R}$ is a topological
invariant, leading thus to empty equations of motion. To second order in 
$b$ we have quartic terms in the extrinsic curvature of the worldvolume.
We can compare this expansion with the one developed for the cosmic string dynamics 
which involves finite width corrections by Anderson {\it et al} \cite{anderson,anderson2}. 
Both expansions possess a slight discrepancy in a piece of the second order 
correction term. 

\subsection{The limit $|b\,f_{ab}|\ll |g_{ab}|$ for the string case}

We survey the application of the formalism developed above by considering 
the string case ($p=1$). For illustrative purposes, we only determine the 
expansions up to first order 
in $b$. We have
\begin{equation*}
 \sqrt{-M} = \sqrt{-g} \left( 1 + \frac{b}{2}\, f 
 \right) ,
\end{equation*}
and
\begin{equation*}
 (M^{-1})^{ab} = g^{ab} - b\,f^{ab} ,
\end{equation*}
where we have used the notation $f = g^{ab} f_{ab}$. 
Putting these results together, it follows immediately that
\begin{equation}
\sqrt{-M} (M^{-1})^{ab} = \sqrt{-g}\left[ g^{ab} 
+ \frac{b}{2}\left( f g^{ab} - 2 f^{ab}\right) 
  \right]
\label{eq:riG}  
\end{equation}
We may therefore, by inserting the expansi\'on~(\ref{eq:riG}) into~(\ref{eq:eom-r}),
infer the equations of motion
\begin{equation*}
 {\cal E}^i _{0} + b\, {\cal E}^i _{1} 
 =0\,,
\end{equation*}
where 
\begin{eqnarray}
 {\cal E}^i _{0} &=&  g^{ab}K_{ab}^i\,,
\label{eq:ee0}
\\
{\cal E}^i _{1} &=& - \widetilde{\Delta} K^i
- \left( f^{ab} - \frac{1}{2} f\,g^{ab} \right) K_{ab} ^i, 
\label{eq:ee1}
\label{eq:ee2}
\end{eqnarray}
where $\widetilde{\Delta} = g^{ab}\widetilde{\nabla}_a \widetilde{\nabla}_b$
is the worldvolume D'Alambertian operator. We identify immediately the DNG and 
Polyakov equations of motion in ${\cal E}^i _0$ and ${\cal E}^i _1$, respectively,
when they vanish\cite{ruth1,ruth2,Noether,defo}. The expressions (\ref{eq:ee1}) result
highly non-trivial. Up to second order in $b$, the associated expansions and the 
ensuing equations of motion are quite involved. We do not pursue this issue 
further here. It would be important to remark that another relevant expansions taking into 
account the string thickness built from the extrinsic curvature in order to 
describe cosmic strings can be found in\cite{anderson,anderson2}.

\section{Born-Infeld-Einstein type brane gravity}

Some time ago, Deser and Gibbons suggested an elegant modification of 
the Einstein-Hilbert (EH) action of general relativity\cite{deser}. 
Their proposal has a determinant Lagrangian density form with the 
spacetime metric and the Ricci tensor adding to form a new metric. Many 
interesting properties like the ghost freedom, regularization of some 
singularities, supersymmetrizability and reduction to EH action at small 
curvatures, made their suggestion very especial indeed. Close in the spirit to the one 
developed by Deser and Gibbons, in the brane context we would like to explore 
such proposal. The action we then consider will be 
\begin{equation}
S_{{\mbox{{\tiny BIE}}}} [X] = \alpha \int_m d^{p+1} \xi \,
\sqrt{-{\mbox{det}}(g_{ab}+ \beta 
\, {\cal R}_{ab})},
\label{eq:bie}
\end{equation}
where $\alpha$ is the tension of the brane, $\beta$ is a concomitant 
dimensional parameter, and ${\cal R}_{ab}$ denotes 
the worldvolume Ricci tensor. Hereafter, we will use the acronym
BIE to denote Born-Infeld-Einstein. ${\cal R}_{ab}$ depends explicitly 
of the extrinsic curvature of the worldvolume, courtesy of the Gauss-Codazzi 
integrability condition\cite{Spivak,defo}. It is given by 
\begin{equation}
 {\cal R}_{ab} = K_i K_{ab} ^i - K_a{}^{c\,i}K_{cb\,i}\,.
\label{eq:ricci}
\end{equation}
Note that the Ricci tensor is symmetric when it is expressed in the 
fashion~(\ref{eq:ricci}). In pure gravity this is not the case in 
general\footnote{Of course, the worldvolume Ricci tensor can be 
expressed also as
${\cal R}_{ab} = 2 \partial_{[c} \Gamma^c _{a]b} - 2 \Gamma^d _{a[c} 
\Gamma^c _{b]d}$.
}. 
In order to introduce auxiliary variables in our description, as before,
we first consider~(\ref{eq:bie}) as a functional with Lagrangian density
$ {\cal L}_{\mbox{\tiny BIE }}= {\cal L}_{{\mbox{{\tiny BIE}}}}
(g_{ab}, K_{ab} ^i)$.
To know the response of the action (\ref{eq:bie}) to a deformation of 
the worldvolume, $X \to X + \delta X$, among auxiliary variables, we 
will continue taking advantage of the results of \ref{app:AVM}, as in the
previous Section.

As before, we will use the notation $(\mathbf{M}^{-1})^{ab}$ to denote 
the inverse matrix of $\mathbf{M}_{ab} 
:= g_{ab} + \beta\,{\cal R}_{ab}$, such that $(\mathbf{M}^{-1})^{ac}
\mathbf{M}_{cb} = \delta^a{}_b$. Expressions (\ref{eq:Tab}) and (\ref{eq:Labi})
specialized to the BIE action result
\begin{eqnarray}
\T^{ab}&=& \alpha \frac{\sqrt{- \mathbf M}}{\sqrt{-g}} (\mathbf{M}^{-1})^{cd}
\left( \delta^a{}_c \delta^b {}_d - \beta\,  
{\cal R}^a{}_c{} ^b{}_d \right),
\label{eq:tt}
\\
\Lambda^{ab} _i &=& -\frac{\alpha \beta}{2}\frac{\sqrt{-\mathbf {M}}}{\sqrt{-g}} 
{\mathbf G}^{abcd} K_{cd\,i}\,,
\label{eq:labi}
\end{eqnarray}
where we have used the worldvolume Riemann tensor
expressed in terms of the extrinsic curvature via the Gauss-Codazzi equation,
${\cal R}_{abcd} = K_{ac} ^i K_{bd\,i} - K_{ad} ^i K_{bc\,i}$, 
and we have introduced the tensor field
\begin{equation}
\mathbf{G}^{abcd} = (\mathbf{M}^{-1})^{ab} g^{cd} + 
(\mathbf{M}^{-1})^{cd} g^{ab} - 2 (\mathbf{M}^{-1})^{c(a} 
g^{b)d}\,.
\label{eq:superMetric}
\end{equation}
The conserved stress tensor for our present case, from Eqs.~(\ref{eq:tt}), (\ref{eq:labi})
and~(\ref{eq:f-final}), is given by
\begin{eqnarray}
\f^a &=& - \left\lbrace   \mathbf{T}^{ab} +   \alpha \beta \frac{\sqrt{-\mathbf M}}{\sqrt{-g}}
\left[ \left( {\mathbf M} ^{-1} \right)^{ac}{\cal R}^b {}_c + \left( {\mathbf M} ^{-1} \right)^{cd}
{\cal R}^a{}_c {}^b{}_c \right] \right\rbrace \,e_b 
\nonumber
\\
&-& \frac{\alpha \beta}{\sqrt{-g}} \widetilde{\nabla}_b \left(\sqrt{-\mathbf M}\,
{\mathbf G}^{abcd}K_{cd\,i}\right) \,n^i\,. 
\end{eqnarray}

As argued earlier, insisting on the conservation law for
$\f^a$, its normal projection produces the equations of motion
\begin{equation}
\T^{ab} K_{ab} ^i = {\mathbf F}^i\,,
\label{eq:eom-Rab}
\end{equation}
where
\begin{equation*}
{\mathbf F}^i = \frac{\alpha \beta}{\sqrt{-g}}\left[ 
\widetilde{\nabla}_a \widetilde{\nabla}_b \left( \sqrt{- \mathbf M}\,
\mathbf{G}^{abcd} K_{cd} ^i\right) - \sqrt{- \mathbf M} 
 \,\mathbf{G}^{abcd} K_{cd} ^j K^e{}_{b\,j} K_{ae} ^i
\right].
\end{equation*}
Once more, apparently the $N-p-1$ equations of motion (\ref{eq:eom-Rab})
resembles Newton's second law, but in general these
equations are of fourth order in derivatives of $X$. There is comparison
only up to first order in $\beta$.  We will back on this point.
Explicitly, the ensuing equations of motion (\ref{eq:eom-Rab}) are
\begin{equation}
\sqrt{-\mathbf{M}}\left[ g^{ab} + (\mathbf{M}^{-1})^{ab}
- \beta (\mathbf{M}^{-1})^{cd} {\cal R}^a{}_c{}^b{}_{d} \right]
K_{ab} ^i 
= \beta \widetilde{\nabla}_a \widetilde{\nabla}_b 
\left( \sqrt{- \mathbf M}\,\mathbf{G}^{abcd} K_{cd} ^i\right) .
\label{eq:eom-Rab-2}
\end{equation}
Once again, these equations of motion are in agreement with those equations of motion
obtained by usual variation methods\cite{Noether,defo}. To my knowledge,
this form of the equations of motion has not been previously discussed. 
By the way, it is possible to exploit the Codazzi-Mainardi equation, 
$\widetilde{\nabla}_a K_{bc} ^i = \widetilde{\nabla}_bK_{ac} ^i$\cite{defo},
in order to separate multiplicatively the extrinsic curvature
on the right-hand side of the previous relation. A straightforward 
computation yields
\begin{eqnarray}
\widetilde{\nabla}_a \widetilde{\nabla}_b 
\left( \sqrt{- \mathbf M}\,\mathbf{G}^{abcd} K_{cd} ^i\right)&=& \left\lbrace  
\nabla^c\nabla_c \left[ \sqrt{-\mathbf{M}} (\mathbf{M}^{-1})^{ab} \right]
+ \nabla_c\nabla_d \left[ \sqrt{-\mathbf{M}} (\mathbf{M}^{-1})^{cd} \right]g^{ab}
\right.
\nonumber
\\
&-& \left. 2 \nabla_c\nabla_d \left[ \sqrt{-\mathbf{M}} (\mathbf{M}^{-1})^{a(c} 
\right] g^{d)b} \right\rbrace K_{ab} ^i.
\label{eq:44}
\end{eqnarray}

For completeness, as in previous cases, working out the expression defining the 
inverse matrix $(\mathbf{M}^{-1})^{ab}$, we have the identity
\begin{equation}
\T^{ab}\mathbf{M}_{bc} = (-g)^{-1/2}{\cal L}_{\mbox{\tiny{BIE}}} \delta^a{}_c -
\beta \mathbf{P}^{ab}{\mathbf{M}}_{bc}\,,
\end{equation}
with $\sqrt{-g}\mathbf{P}^{ab}:= \alpha \,\sqrt{-\mathbf{M}}
(\mathbf{M}^{-1})^{cd} {\cal R}^a{}_c{}^b{}_d $.

\subsection{Alternative geometry}

Analogous to the foregoing Section, for this model we can think also of an alternative volume 
element, now depending of the worldvolume metric and a term quadratic in the
extrinsic curvature by the Ricci tensor. 
Suppose now that $\mathfrak{D}_a$ is the covariant derivative compatible 
with $\mathbf{M}_{ab}= g_{ab} + \beta {\cal R}_{ab}$. Then, according to this
new geometry the corresponding connection coefficients $C^a _{bc}$,
are given by
\begin{equation}
C^a _{bc} = \Gamma^a _{bc} + \frac{\beta}{2} ({\mathbf{M}}^{-1})^{ad} \left( \nabla_b 
{\cal R}_{cd} + \nabla_c {\cal R}_{bd} - \nabla_d {\cal R}_{bc}\right) ,
\label{eq:Css}
\end{equation}
where $\Gamma^a _{bc}$ are the connection coefficients associated
with the induced metric $g_{ab}$. The corresponding relationship between the
Ricci tensors is now the following
\begin{equation}
{\cal R}_{ab} = {\mathbf R}_{ab} + {\mathfrak D}_a {\mathbf S}^c _{bc} - 
{\mathfrak D}_c {\mathbf S}^c _{ab} - {\mathbf S}^d _{ac} {\mathbf S}^c _{bd} + 
{\mathbf S}^d _{ab} {\mathbf S}^c _{cd}\,,
\end{equation}
where $ {\mathbf R}_{ab}$ is the Ricci tensor associated with the
connection $C^a _{bc}$ and we have defined
\begin{equation}
 {\mathbf S}^a _{bc} := \frac{\beta}{2} ({\mathbf M}^{-1})^{ad} \left( \nabla_b 
{\cal R}_{cd} + \nabla_c {\cal R}_{bd} - \nabla_d {\cal R}_{bc}\right) . 
\end{equation}
By considering this new structure it is straightforward to obtain the identity
$\nabla_a \sqrt{-\mathbf{M}}= \sqrt{-\mathbf{M}}\, {\mathbf S}^b _{ab} $.

As before we turn now to expand the integrand of the action~(\ref{eq:action}) using 
the expression~(\ref{eq:expansion}). It is straightforward to note that
\begin{equation}
\sqrt{-\mathbf{M}} = \sqrt{-g} \left\lbrace 1 + \frac{\beta}{2}\,
{\cal R} + \frac{\beta^2}{8} \left( {\cal R}^2 - 2{\cal R}_{ab} {\cal R}^{ab}  
\right) +{\cal O}({\cal R}^3)  \right\rbrace .
\end{equation}
We observe that to first order in $\beta$ we have the sum of the DNG model and 
the RT model\cite{RT,ostro}. Here it is worthy to note that
the case of a string ($p=1$) is quite special since the term $\sqrt{-g}{\cal R}$ 
corresponds to a total derivative and in consequence we have a model classically
equivalent to the DNG action. Furthermore, in general to first order in $\beta$ 
the equations of motion results of second order in the derivatives of the fields. 
Note that only up to second order in $\beta$, the strong second order derivative 
terms come into play where we have quartic terms in the extrinsic curvature of the 
worldvolume.

\subsection{The limit $|\beta \,{\cal R}_{ab}| \ll |g_{ab}|$ for the
brane world scenarios}

A special case worth pointing out is the $p=3$ case. From the expansions 
up to second order in $\beta$ of the main geometrical components given by
\begin{equation*}
 \sqrt{-{\mathbf M}} = \sqrt{-g} \left[ 1 + \frac{\beta}{2} {\cal R} 
+ \frac{\beta^2}{8} \left({\cal R}^2 - 2{\cal R}_{ab} {\cal R}^{ab} \right) 
\right] \,,
\end{equation*}
and
\begin{equation*}
 ({\mathbf M}^{-1})^{ab} = g^{ab} - \beta \,{\cal R}^{ab}
+ \beta^2 \, {\cal R}^a{}_c {\cal R}^{cb}\,,
\end{equation*}
we evidently get
\begin{eqnarray*}
\sqrt{-{\mathbf M}} ({\mathbf M}^{-1})^{ab} &=&
\sqrt{-g}\left\lbrace g^{ab} - \beta\, G^{ab} + \frac{\beta^2}{8}
\left[ \left( {\cal R}^2 - 2 {\cal R}_{cd}{\cal R}^{cd}
\right)g^{ab} - 4{\cal R}{\cal R}^{ab} 
\right. \right.
\\
&+& \left. \left. 8 {\cal R}^a {}_c {\cal R}^{cb}
 \right]   \right\rbrace .
\end{eqnarray*}
where $G_{ab}={\cal R}_{ab} - \frac{1}{2}{\cal R}\,g_{ab}$ is the worldvolume
Einstein tensor.

Therefore, taking into account the expression (\ref{eq:44}) and after
some algebra, the equations of motion coming from (\ref{eq:eom-Rab-2}) 
can be rearranged into
\begin{equation}
 {\cal E}^i _{0} + \beta\, {\cal E}^i _{1} + \beta^2\,{\cal E}^i _{2} =0\,,
 \label{eq:eofmoto}
\end{equation}
where
\begin{eqnarray}
 {\cal E}^i _{0} &=&  g^{ab} K_{ab}^i\,,
\label{eq:e0}
\\
{\cal E}^i _{1} &=& - G^{ab}K_{ab} ^i,
\label{eq:e1}
\\
{\cal E}^i _{2} &=& 
 S^{ab} K_{ab} ^i
\label{eq:e2}
\end{eqnarray}
where we have introduced the tensor field
\begin{equation}
S_{ab} = \nabla^c \nabla_c {\cal R}_{ab} - \frac{1}{2}g_{ab}\nabla^c \nabla_c 
{\cal R} + 2{\cal R}^{cd}{\cal R}_{acbd} -  {\cal R}{\cal R}_{ab} +
\frac{1}{4}g_{ab}\left( {\cal R}^2 - 2 {\cal R}_{cd}{\cal R}^{cd}\right) .
\label{eq:Sab}
\end{equation}
It must be noted that up to second order in $\beta$, the fourth-order equations 
of motion in the field variables appear. These equations are geometrical 
and physically correct. The $N-p-1$ equations~(\ref{eq:e0}), if vanish, 
correspond to the well known DNG equations of motion\cite{defo}. Equally,
when the expressions~(\ref{eq:e1}) vanish, these correspond to RT equations of 
motion\cite{RT,ostro,hamRT}. Notice that, the field equations for the string case 
($p=1$) are those of the DNG case. Moreover, the remaining 
relations~(\ref{eq:e2}) are quite involved. These are  
highly non-linear field equations. Such equations of motion are 
obtained from the model $L={\cal R}^2 - 2{\cal R}_{ab} {\cal R}^{ab}$
which has recently proposed as a brane correction to the Dvali-Gabadadze-Porrati
model proposed as a solution for the cosmological constant problem\cite{cadoni}.
Actually, the form as the equations of motion (\ref{eq:eofmoto}) are presented
allows us to identify the piece associated to the $\beta^2$ expression as a Weyl term\cite{bostock}
which is a geometric correction to the RT equations of motion\footnote{Likewise
to the Einstein tensor in the case that the field variable is the induced metric.}.
The dynamics of this type of extended objects is highly non-trivial
but striking in the search of a proper solution of the cosmological
constant problem. This model require a more careful and deep study under 
this approach which will be reported elsewhere.

\section{Concluding remarks}

A simple recipe has been presented for studying the dynamical structure 
of some BI type brane theories by exploiting the mechanical content 
that the conserved stress tensor possesses. This is achieved 
by using an auxiliary variables method. After considering appropiate quadratic
geometrical constraints it is possible to obtain effortless the response of 
every considered BI type brane action to a deformation $X \to X + 
\delta X$ once it is distributed among the auxiliary variables. We have 
seen that judicious choices for the constraints are intended to facilitate 
the calculation of the brane dynamics thus providing an economical and
efficient way to explore the motion of interesting extended objects. In 
addition, we have presented a Born-Infeld-Einstein action adapted to the
brane context, closed in spirit of the one developed by Deser and 
Gibbons\cite{deser}. To my knowledge, this action has not been previously
discussed in the brane content.
This action contains a rich geometrical content
and add new insights with likely benefit for the study of branes
in a cosmological context. 
To lowest order in the rigidity parameter, it contains 
both DNG and RT models, which are first order theories. Nevertheless, 
the full second order derivative dependence is switched on up to second 
order terms in the rigidity term expansion. This modify substantially the 
RT gravity\cite{RT,ostro} by incorporating quadratic terms in the Ricci curvature. 
It would be interesting to investigate the relevance of 
this model in the brane world universes context where interesting
physical background spacetimes must be considered. It is intended to discuss this
in a separate work.

\section*{Acknowledgments}

I thank Alberto Molgado for his constructive criticism 
and fruitful discussions. I also thank E. Ay\'on-Beato 
for encouragement and valuable comments. This work was partially 
supported from grants PROMEP-UV (CA Investig. y Ense\~nanza de la F\'isica)
and PROMEP-UAZ-PTC-086.

\appendix

\section{Auxiliary variables method}
\label{app:AVM}

Consider a brane, denoted by $\Sigma$, of dimension $p$ evolving in a
$N$-dimensional fixed Minkowski background spacetime with metric
$\eta_{\mu \nu}$\quad $(\mu\,\nu = 0,1,\ldots, N-1)$.
The worldvolume $m$, swept out by $\Sigma$, is a oriented timelike
manifold of dimension $p+1$, described by the embedding functions
$x^\mu = X^\mu (\xi^a)$ where $x^\mu$ are local coordinates of the
background spacetime, $\xi^a$ are local coordinates of $m$, and $X^\mu$
represent the embedding functions $(a,b=0,1,\ldots,p)$. The metric 
induced and the extrinsic curvature defined on the worldvolume are 
\begin{equation}
g_{ab} = e_a \cdot e_b \qquad K_{ab} ^i = e_a \cdot \widetilde{\nabla}_b
n^i \,,
\label{eq:tensors}
\end{equation}
where $\widetilde{\nabla}_a$ is the $O(N-p-1)$ covariant derivative
on $m$ and also invariant under normal rotations
and we have exploited the Gauss-Weingarten equation, $\widetilde{\nabla}_a
n^i = K_{ab}^i g^{bc}e_c$\cite{defo}. Here, $e_a=\partial_a X$ and $n^i$ 
denote the tangent and unit normal vectors defined on $m$, respectively
\begin{equation}
n^i \cdot e_a = 0 \qquad n^i \cdot n^j = \delta^{ij},
\label{eq:normal}
\end{equation}
where $(i,j= 1,2,\ldots, N-p-1)$. Note that the tensors (\ref{eq:tensors})
encode derivatives of the field variables $X$.
Suppose that the following generic action governs the dynamics
of $\Sigma$
\begin{equation}
S_0[X] = \int_m d^{p+1}\xi\,{\cal L}(g_{ab}, K_{ab} ^i)\,,
\label{eq:SS}
\end{equation}
where ${\cal L}= {\cal L}(g_{ab},K_{ab}^i)$ is the Lagrangian density
of the field theory underlaying the dynamics of $\Sigma$.
At the technical level, this type of Lagrangians are complicated
to handle.

In order to know the response of the action (\ref{eq:SS}) to a 
deformation of the worldvolume, $X \to X + \delta X$, among 
auxiliary variables, we must first construct an extended action
by considering constraints manifestly\cite{Guven},
\begin{eqnarray}
S &&\!\!\!\!\!\!\![ X,e{}_a,n^i,g_{ab},K_{ab} ^i , f^a ,\Lambda^{ab} _i ,\Lambda^{ab},\Lambda_{ij},
\Lambda^a{}_i]=  S_0 [g_{ab},K_{ab} ^i ] + \int_m dV\,f^a 
\cdot (e {}_a - \partial_a X) 
\nonumber \\
&+& 
\int_m dV\,\Lambda^a{}_i \,(e{}_a \cdot n^i) +
\frac{1}{2} \int_m dV\,\Lambda_{ij}
\,(n^i \cdot n^j - \delta^{ij})   - \frac{1}{2}\int_m
dV\, T^{ab}(g_{ab} - e{}_a \cdot e{}_b) \nonumber
\\ 
&+&  \int_m dV\,\Lambda^{ab} _i 
\left( K_{ab} ^i   - e_a \cdot \widetilde{\nabla}_b n^i \right).
\label{eq:S-total}
\end{eqnarray}
In this approach, $f^a,\Lambda^a {}_i, \Lambda_{ij},T^{ab}$ and $ \Lambda^{ab} _i$ are 
Lagrange multipliers enforcing the definitions of the auxiliary variables
(\ref{eq:tensors}) and (\ref{eq:normal})
whereas $X,e_a,n^i,g_{ab}$ and $K_{ab} ^i$ are considered as
independent fields. We perform the variation in steps. 
The variation of~(\ref{eq:S-total}) with respect to $X$ reproduces 
the covariant conservation of  $f^a$
\begin{equation}
\nabla_a f^a = 0.
\label{eq:CC} 
\end{equation}
Moreover, varying  over tangent vectors we get that
\begin{equation}
f^{a}= - \left( T^{ab} -  \Lambda^{ac} _i K_c {}^{b\,i}  \right) e_b - 
\Lambda^a {}_i \,n^i \,
\label{eq:faa}
\end{equation}
results an expression for $f^a$ as a linear expansion in terms
of the worldvolume basis. This occurs since we are dealing with a 
second order derivative theory.
The EL derivative of the action~(\ref{eq:S-total}) with respect 
to the induced metric casts out
\begin{equation}
T^{ab}= \frac{2}{\sqrt{-g}}\left( \frac{\partial {\cal L}}{\partial g_{ab}}
\right).
\label{eq:Tab}
\end{equation}
This correspond to the worldvolume stress tensor.
Similarly, the EL derivative with respect to the extrinsic curvature results
\begin{equation}
\Lambda^{ab} _i= - \frac{1}{\sqrt{-g}} \left( \frac{\partial {\cal L}}{\partial K_{ab} ^i}\right).
\label{eq:Labi}
\end{equation}

Moreover, the geometrical information contained in the Lagrange multipliers 
is associated with the EL derivative with respect to the normals $n^i$.
They yield 
\begin{eqnarray}
 \Lambda^a{}_i &=& - \widetilde{\nabla}_b \Lambda^{ab} _i ,
\\
\Lambda_{ij} &=& \Lambda^{ab} _{(i} K_{|ab|j)},
\end{eqnarray}
where we have considered, $\nabla_a e_b = - K_{ab} ^i n_i$.
Finally, the conserved stress tensor is given by
\begin{equation}
f^a = - \left( {T}^{ab} - \Lambda^{ac} _j K_c{}^{b\,j}
\right)\,e_b + \left( \widetilde{\nabla}_b \Lambda^{ab} _i
\right)\,n^i\,. 
\label{eq:f-final} 
\end{equation}

As argued earlier, in order to explore the contents of the conservation 
law for $f^a$, we observe that its normal 
projection produces the equations of motion, usually written as
\begin{equation}
T^{ab} K_{ab} ^i = {F}^i\,,
\end{equation}
with
\begin{equation*}
{F}_i = - \widetilde{\nabla}_a \widetilde{\nabla}_b \Lambda^{ab} _i
+ \Lambda^{ac} _j K_c{}^{b\,j} K_{ab\,i}
\end{equation*}
whereas the tangential part is given by
\begin{equation}
\nabla_a T^{ab} =  \Lambda^{ac} _i \widetilde{\nabla}_a K_c{}^{b\,i}
+ 2 \widetilde{\nabla}_a \Lambda^{ac} _j \,K_c{}^{b\,j}.
\label{eq:tangential}
\end{equation}
These latter are consistency conditions 
which are consequence of the reparametrization invariance of the 
action (\ref{eq:SS}).
Apparently the $N-p-1$ equations of motion (\ref{eq:eom-Rab})
resembles Newton's second law, but in general these
equations are of fourth order in derivatives of $X$.

\section{Determinant of a matrix}
\label{app:mathidentities}

The Levi-Civita tensor in $n$-dimensions, $\epsilon_{a_1,a_2, \ldots,a_n}$,
is related to a totally antisymmetric pseudotensor by the relation
\begin{equation}
\epsilon_{a_1,a_2, \ldots,a_n} = \sqrt{-g}\varepsilon_{a_1,a_2, \ldots,a_n}\,,
\end{equation}
where $\varepsilon_{a_1,a_2, \ldots,a_n}$ is the Levi-Civita pseudotensor
which is a tensorial density of weight $\omega = -1$.

\noindent
The determinant of a matrix $M_{ab}$ can be defined in terms of the
Levi-Civita pseudotensor by
\begin{equation}
M := {\mbox{det} (M_{ab})} = \frac{1}{n!} \varepsilon^{a_1,a_2, \ldots,a_n}
\varepsilon^{b_1,b_2, \ldots,b_n}M_{a_1 b_1} M_{a_2 b_2}\ldots M_{a_n b_n},
\end{equation}
where we assume that $\varepsilon_{1,2,\ldots,n}=1$. When $M_{ab}$
is non-singular, its inverse matrix has a representation in terms
of the Levi-Civita pseudotensor
\begin{equation}
\left( M^{-1} \right)^{a_1 b_1} = \frac{1}{(n-1)! M}
 \varepsilon^{a_1,a_2,a_3, \ldots,a_n}
\varepsilon^{b_1,b_2,b_3, \ldots,b_n} M_{b_2 a_2} M_{b_3 a_3} \ldots
M_{b_n a_n}.
\end{equation}

\vspace{0.4cm}

\noindent
Let $A$ be a $n \times n$ matrix. It results useful to expand the 
determinant of $M := I + a \,A$,
\begin{equation}
\left[\, \mbox{det}\, ( I + a\,A ) \, \right]^{1/2} = 1 + \frac{a}{2}
\,\mbox{Tr}\,A + \frac{a^2}{8}\left[ \left( \mbox{Tr}\,A \right)^2
- 2 \mbox{Tr}\, A^2 \right] + {\cal O}\,(A^3)\,,  
\label{eq:expansion}
\end{equation}
where $a$ is a constant and $I$ being the $n\times n$ identity matrix.



\begin{thebibliography}{0}    


\bibitem{Born} M. Born, {\it Phys. Roy. Soc.} {\bf A143} (1934) 410.

\bibitem{Born1} M. Born and L. Infeld, {\it Phys. Roy. Soc.}
{\bf A144} (1934) 425.

\bibitem{Polchinski1} J. Polchinski, {\it Phys. Rev. Lett.} {\bf 75} (1995) 4724.

\bibitem{clifford} C. Johnson {\it D-branes} (Cambridge University Press, 
Cambridge, UK, 2003).

\bibitem{Lindstrom2} U. Lindstrom, M. Rocek and P. Van Nieuwenhuizen, 
{\it Phys. Lett. B} {\bf 199} (1987) 219; {\it Phys. Lett. B} 
{\bf 201} (1988) 63.

\bibitem{Lindstrom} U. Lindstrom, {\it Int. J. Mod. Phys. A}
{\bf 3} (1988) 2401.

\bibitem{brandon-ruth} B. Carter and R. Gregory, {\it Phys. Rev. D} {\bf 51}
 (1995) 5839.

\bibitem{anderson} M. Anderson, F. Bonjour, R. Gregory and J. Stewart, {\it Phys. Rev. D}
{\bf 56} (1997) 8014.

\bibitem{anderson2} M. Anderson, {\it Phys. Rev. D}{\bf 51} 2863 (1995).

\bibitem{polyakov} A. M. Polyakov, {\it Nucl. Phys. B} {\bf 268} (1986) 406.

\bibitem{kleinert} H. Kleinert, {\it Phys. Lett. B} {\bf 174} 335 (1986).

\bibitem{ruth1} D. Garfinkle and R. Gregory, {\it Phys. Rev. D} {\bf 41} 1889 (1990).

\bibitem{ruth2} R. Gregory, {\it Phys. Rev. D} {\bf 43} 520 (1991)

\bibitem{ruth3} B. Carter and R. Gregory, {\it Phys. Rev. D} {\bf 51} 5839 (1995).
 
\bibitem{randall} L. Randall and R. Sundrum, {\it Phys. Rev. Lett.} {\bf 83} 3370
(1999); {\bf 83} 4690 (1999).

\bibitem{cadoni} M. Cadoni and P. Pani, {\it Phys. Lett. B} {\bf 674} (2009) 308.

\bibitem{Noether} G. Arreaga, R. Capovilla and J. Guven,
{\it Ann. Phys. NY} {\bf 279} (2000) 126.

\bibitem{Guven} J. Guven, {\it J. Phys. A:Math. and Theor.} {\bf 37} (2004) 
L313.

\bibitem{dbigeometry} R. Cordero, A. Molgado and E. Rojas,
{\it Class. Quant. Grav.} {\bf 24} (2007) 1665.

\bibitem{deser} S. Deser and G. W. Gibbons, {\it Class. Quant. Grav.}
{\bf 15} (1998) L35-L39.

\bibitem{Spivak} M. Spivak, {\it Comprehensive Introduction to Differential Geometry}
Vol. 4 2nd edn (Publish or Perish, Boston, MA, 1970).

\bibitem{defo} R. Capovilla and J. Guven, {\it Phys. Rev. D} {\bf D51} (1995) 6736.

\bibitem{Carter2} B. Carter, {\it Int. J. Theor. Phys.} {\bf 40} (2001) 2099.

\bibitem{bostock} P. Bostock, R. Gregory, I. Navarro and J. Santiago, {\it Phys. Rev. Lett.}
{\bf 92} (2004) 221601

\bibitem{wald} R. Wald, {\it General Relativity} (The University of Chicago
Press, 1986).

\bibitem{RT} T. Regge  and C. Teitelboim, Gravity \`a la string, in 
{\it Proceedings of the Marcel Grossman Meeting, Trieste, Italy 1975}, eds. R. Ruffini 
(North-Holland, Amsterdam, 1977), p.~77.

\bibitem{ostro} R. Cordero, A. Molgado and E. Rojas, {\it Phys. Rev. D} {\bf 79}
(2009) 024024.

\bibitem{hamRT} R. Capovilla, A. Escalante, J. Guven and E. Rojas,
hep-th/0605160. 

\end{thebibliography}
\end{document}